\documentclass[fleqn,usenatbib]{mnras}

\usepackage[T1]{fontenc}
\usepackage{ae,aecompl}

\usepackage{graphicx}	
\usepackage{amsmath}	
\usepackage{amssymb}	

\title[Evolution of SFR-density relation]{
Evolution of star formation rate-density relation over cosmic time in a simulated universe: 
the observed reversal reproduced}

\author[H. S. Hwang, J. Shin and H. Song]{
Ho Seong Hwang$^{1,2}$\thanks{E-mail: hhwang@kasi.re.kr},
Jihye Shin$^{1,2}$\thanks{E-mail: jhshin@kasi.re.kr} and
Hyunmi Song$^{1,2}$\thanks{E-mail: hmsong@kasi.re.kr}
\\
$^{1}$Korea Astronomy and Space Science Institute, 
776 Daedeokdae-ro, Yuseong-gu, Daejeon 34055, Korea\\
$^{2}$School of Physics, Korea Institute for Advanced Study,
85 Hoegiro, Dongdaemun-gu, Seoul 02455, Korea
}

\date{Accepted XXX. Received YYY; in original form ZZZ}

\pubyear{2017}

\begin{document}
\maketitle

\begin{abstract}
We use the IllustrisTNG cosmological hydrodynamical simulation
  to study the evolution of star formation rate (SFR)-density relation over cosmic time.
We construct several samples of galaxies at different redshifts from $z=2.0$ to $z=0.0$,
  which have the same comoving number density.
The SFR of galaxies decreases with local density
  at $z=0.0$, but its dependence on local density becomes weaker with redshift.
At $z\gtrsim1.0$, the SFR of galaxies increases with local density
 (reversal of the SFR-density relation),
  and its dependence becomes stronger with redshift.
This change of SFR-density relation with redshift still remains 
  even when fixing the stellar masses of galaxies.
The dependence of SFR on the distance to a galaxy cluster also shows
	a change with redshift in a way similar to the case based on local density,
	but the reversal happens at a higher redshift, $z\sim1.5$, in clusters.
On the other hand, the molecular gas fraction always 
  decreases with local density regardless of redshift at $z=0.0-2.0$
  even though the dependence becomes weaker when we fix the stellar mass.
Our study demonstrates that the observed 
  reversal of the SFR-density relation at $z\gtrsim1.0$ can be successfully reproduced
  in cosmological simulations.
Our results are consistent with the idea that
  massive, star-forming galaxies are strongly clustered at high redshifts,
  forming larger structures. These galaxies then consume their gas
  faster than those in low-density regions through frequent interactions with other galaxies,
  ending up being quiescent in the local universe.
\end{abstract}



\section{INTRODUCTION}

Galaxy environments strongly affect the physical properties of galaxies through various physical
  mechanisms \citep{bg06,ph09}.
These environmental effects result in tight correlations between galaxy properties and environment: 
  e.g. morphology-density relation (e.g. \citealt{dre80,pg84}), 
  star formation rate (SFR)-density relation (e.g. \citealt{bal98,lew02,gom03}), 
  colour-density relation (e.g. \citealt{hogg03,bla05cdr}).

These relations have been extensively studied in the local universe \citep{bm09}, 
  and have been explored at high redshifts using several deep field survey data
  (e.g. \citealt{cuc06,coo07,hp09,gro18,pel19}).
Understanding this environmental dependence of galaxy properties and 
  its evolution with redshift is very important in modern astrophysics 
  because it provides strong constraints 
  on the model of galaxy formation and evolution \citep{peng10}.
Moreover, thanks to recent cosmological hydrodynamical simulations that 
  have implemented diverse baryonic physics \citep{dub14, vog14, sch15, kha15, pil18tng},  
  we can directly test the predictions from simulations with the results from observations.

Among many correlations between galaxy properties and environment,
  the SFR-density relation has been received particular attention 
  because of its dramatic change with redshift;
  the spatially averaged SFR of galaxies decreases 
  with local density at low redshifts,
  but increases with local density at high redshifts (i.e. $z\sim1.0$).
This is known as  the reversal of SFR-density relation \citep{elb07, coo08}.

Similarly, the average SFRs of cluster galaxies
  are lower than those of field galaxies in the local universe (e.g. \citealt{bg06,ph09, von10}).
Then, the SFR increases with redshift for 
  both cluster and field galaxies,
  but cluster galaxies show much faster increase of SFR than field galaxies.
In the result, the SFRs of cluster galaxies, on average,
  are higher than those of 
  field galaxies at high redshifts (e.g. $z\gtrsim1.3$ in \citealt{alb14, tran10}).
Although the difference in SFR 
  between cluster and field galaxies at high redshifts
  can vary depending on cluster sample,
  some studies also found that
  the SFRs of cluster galaxies are, on average, not lower than 
  those of field galaxies (e.g. \citealt{bro13,alb16}).
  
The reversal of SFR-density relation 
  was initially claimed by \citet{elb07} using the data 
  from the GOODS deep field \citep{gia04}, 
  and has been examined in other deep fields even though 
  some could not reproduce exactly the same trend (e.g. \citealt{sco13,zip14}).
Interestingly, there were some attempts to reproduce the evolution
  of SFR-density relation using numerical simulations, but
  many of them failed to reproduce such a reversal (e.g. \citealt{elb07, sco13}).
  
To our knowledge, \citet{tc14} is the only study that succeeded in reproducing
  the reversal of SFR-density relation 
  using hydrodynamical simulations.
They could successfully reproduce the observed SFR- and 
  colour-density relations at $z=0$ and $z=1$
  even though their analysis is limited to only the two epochs and
  the box size is rather small (i.e. $21\times24\times20~h^{-3}$ Mpc$^3$).
  
Here, we use a recent cosmological hydrodynamical simulation with large volumes 
  to study the evolution of SFR-density relation in more detail.
Among many recent cosmological simulations
  with baryonic physics (e.g. Horizon-AGN: \citealt{dub14},
  Illustris: \citealt{vog14}, EAGLE: \citealt{sch15}
  MassiveBlack-II: \citealt{kha15}),
  we analyse the most recent one, the IllustrisTNG \citep{pil18tng}.
We focus on the evolution of SFR-density relation with redshift
  by analysing the simulation data in a way similar 
  to the observational data.
We describe the simulation data we use in Section \ref{data}, and
 examine the evolution of SFR-density relation in Section \ref{results}.
We  discuss and summarize the results in Sections \ref{discuss} and \ref{sum}, respectively.
Throughout,
  we adopt cosmological parameters as in the TNG simulation: 
  $\Omega_{\Lambda,0}=0.6911$, $\Omega_{m,0}=0.3089$,
  $\Omega_{b,0}=0.0486$, $\sigma_8=0.8159$ 
  and $n_s=0.9667$ and $h = 0.6774$ \citep{planck16}.
All quoted errors in measured quantities are 1$\sigma$, and 
  spatial quantities and coordinates are expressed in comoving units.

\begin{figure*}
	\includegraphics[width=0.85\textwidth]{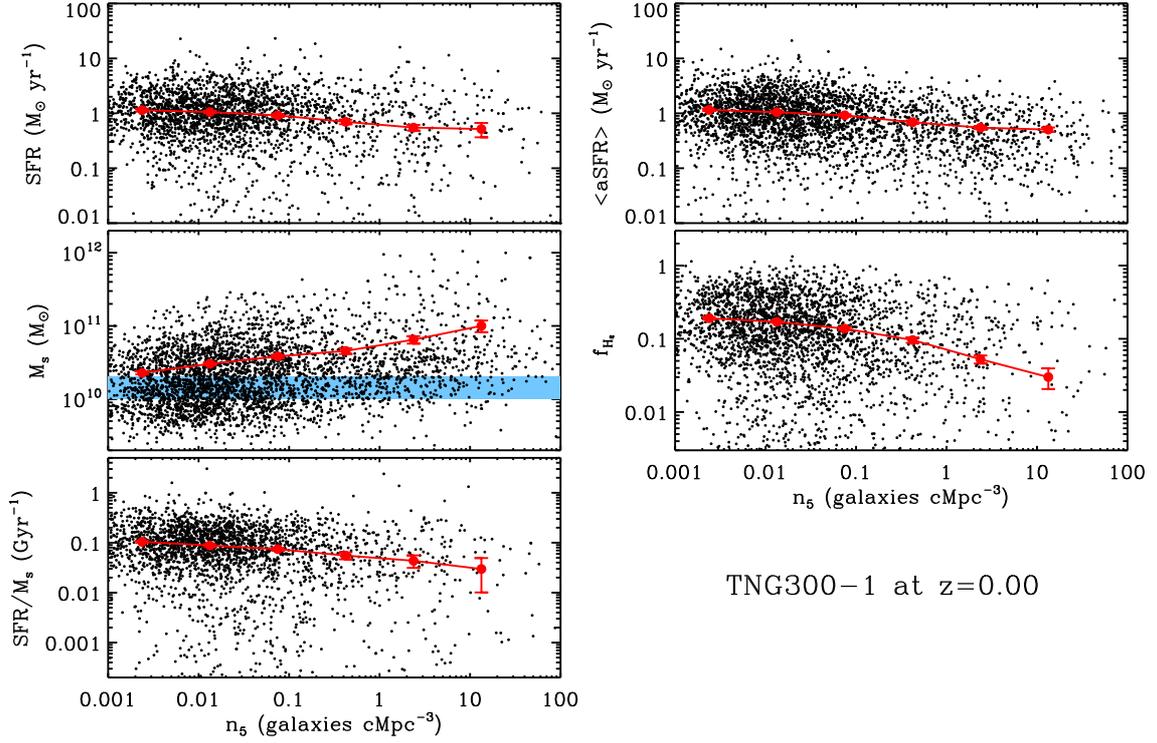}
	\caption{Physical parameters of galaxies in the IllustrisTNG 
		as a function of local density ($n_5$) at $z=0.0$. 
		Black points are individual galaxies, and 
		red points with lines are means of black points.
		The errorbars are derived from the resampling method.
		(top left) star formation rate, 
		(middle left) stellar mass, 
		(bottom left) specific star formation rate,
		(top right) spatially averaged star formation rate, 
		(middle right) mass fraction of molecular hydrogen gas ($f_{\rm H_2} = M_{\rm H_2}/M_{\rm star}$).
		We display only 3\% of the data for clarity.
	}
	\label{z0}
\end{figure*}

\begin{figure*}
	\includegraphics[width=0.85\textwidth]{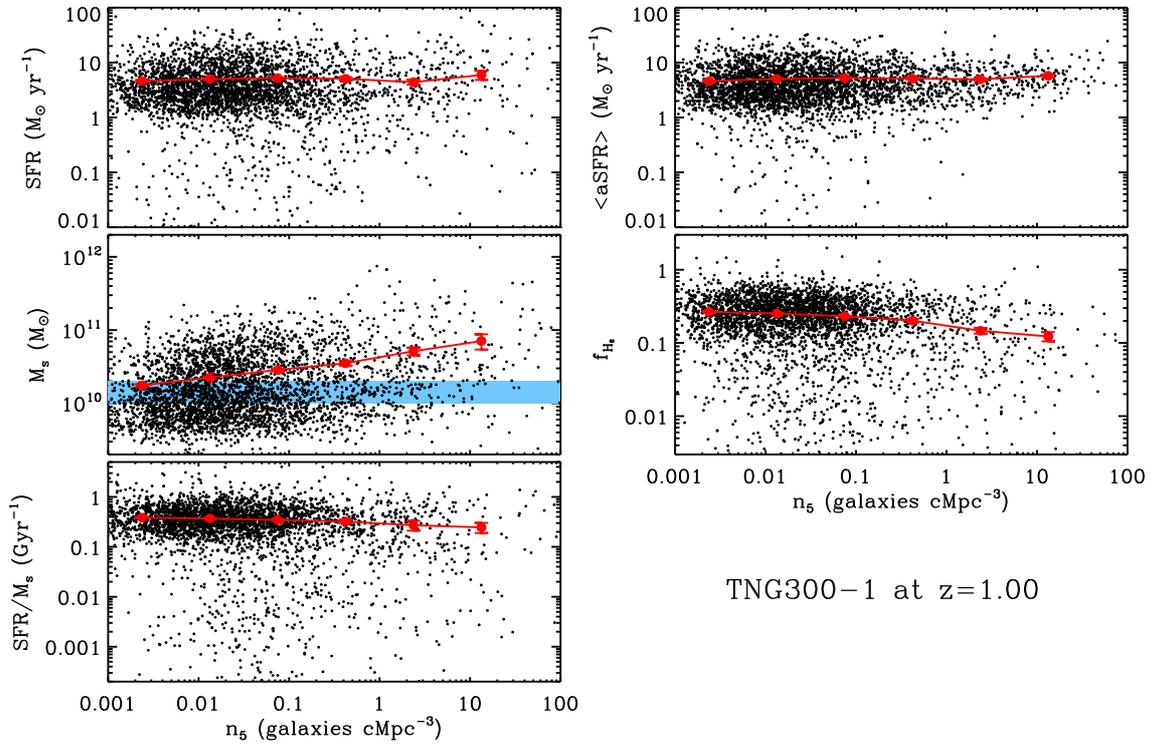}
	\caption{Similar to Fig. \ref{z0}, but for $z=1.0$.}
	\label{z1}
\end{figure*}

\section{Data}\label{data}

The IllustrisTNG is a suite of cosmological hydrodynamic simulations \citep{pil18tng}, 
  which is composed of three simulations 
  with different volumes: TNG300, TNG100 and TNG50.
Each number refers to the length of a side in the simulation box,
  corresponding to roughly 300, 100, and 50 cMpc.
For each simulation box, there are three types of simulations 
  with different mass/spatial resolutions of dark matter particles/gas cells.
For each resolution, there are two types of simulations
  depending on the inclusion/exclusion of gas dynamics 
  (i.e. dark matter only vs. hydrodynamical).
The simulation is run with the moving-mesh code AREPO \citep{spr10},
  and starts from a high redshift at $z=127$ (age$_{\rm universe}=11.44$ Myr) 
  to the current epoch at $z=0$ (age$_{\rm universe}=13.80$ Gyr).

The TNG is the successor of the Illustris simulation \citep{vog14},
  and has improved in many aspects.
For example, the largest simulation box in TNG is 300 cMpc wide,
  which is larger than before (100 cMpc); 
  this allows us to study 
  large-scale galaxy clustering with small cosmic variance,
  and to identify rare, massive galaxy clusters. 
A new effective AGN feedback implementation seems to work better
  in reproducing the observed trends of galaxies
  such as colour-magnitude diagrams \citep{nel18}.
A cosmic magnetism is also considered.
More detailed information is available in their presentation papers
  \citep{pil18,spr18,nel18,nai18,mar18}.

We use the data from the TNG300 (largest volume in the TNG simulation),
   which was publicly released in 2018 \citep{nel19}.
Among several versions of the TNG300 simulations, we use the one
  that has the highest mass resolution with gas dynamics (i.e. TNG300-1).
The number of dark matter particles is 2500$^3$ 
  in a simulation box of (302.6 cMpc)$^3$.
The mass of dark matter particles is $5.9\times10^7$ M$_\odot$.
The TNG provides 100 snapshot data from $z=20.05$ (age$_{\rm universe}=0.18$ Gyr)
  to $z=0.0$ (age$_{\rm universe}=13.80$ Gyr).

The galaxy catalogue we use\footnote{\url{http://www.tng-project.org/data/docs/specifications/}}
  is the output from the SUBFIND subhalo finding algorithm \citep{spr01}.
Among several physical parameters of galaxies available in this catalogue,
  those we mainly adopt in this study are
  star formation rate, stellar mass, $r$-band absolute magnitude and molecular hydrogen mass.
The star formation rate is the sum of the individual star formation rates 
  of all gas cells in a subhalo.
The stellar mass is the total mass of all member star particles
   bound to a subhalo.
The $r$-band absolute magnitude is the total magnitude of member star particles,
  which is modelled with stellar population models \citep{nel18}.
The molecular hydrogen mass is adopted from \citet{die18, die19}.
Among five hydrogen mass measurements based on different models,
  we adopt the one based on \citet{gd14}; the use of different models
  does not change our conclusions.

\begin{figure*}
	\centering
	\includegraphics[scale=0.9]{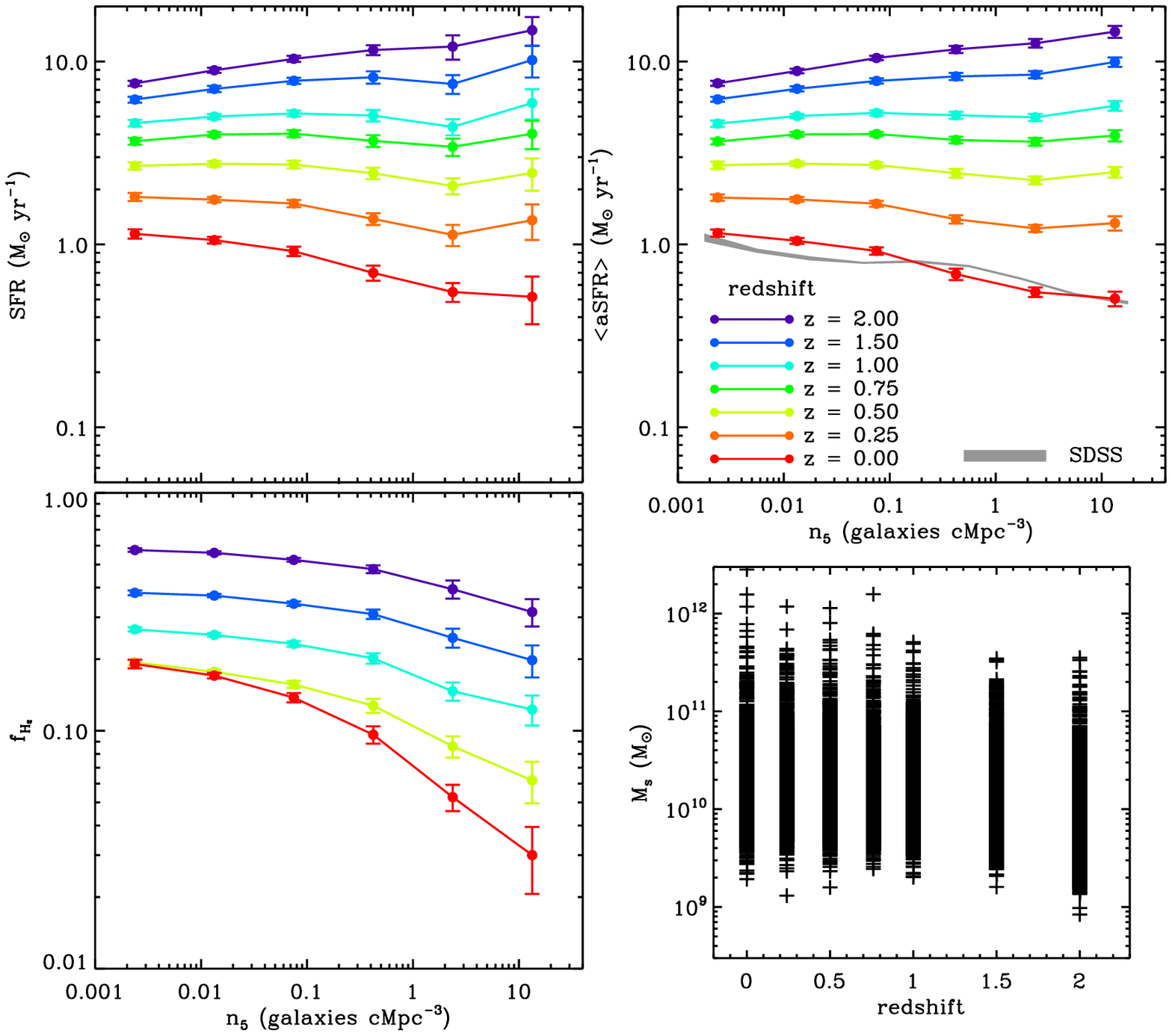}
	\caption{SFRs (top left), average SFRs (top right) and mass fraction of molecular hydrogen (bottom left)
       of galaxies as a function of local density ($n_5$)
		for different redshifts from $z=2.0$ to $z=0.0$.
		The grey line is for the SDSS galaxies in the local universe (Fig. 17 in \citealt{hwa10lirg}).
		The molecular hydrogen masses are available only for the redshifts at $z=0.0$, 0.5, 1.0, 1.5 and 2.0.
		(bottom right) Stellar mass distribution of the galaxies in our samples.}
	\label{asfr}
\end{figure*}

To make fair comparisons between different redshifts,
  at each redshift 
  we make a sample of galaxies with the same comoving number density.
We choose the galaxy number density, 
  corresponding to the one used for the observational data
  of the Sloan Digital Sky Survey (SDSS; \citealt{york00}):
  e.g. 4.43 galaxies cMpc$^{-3}$ as in \citet{hwa10lirg}. 
Therefore, there are 122,764 galaxies 
  in the TNG simulation box of (302.6 cMpc)$^3$ at each redshift.
To make a galaxy sample similar to the magnitude-limited samples
  as in many observations,
  we first sort the galaxies in a given snapshot 
  according to their $r$-band absolute magnitudes.
We then select top 122,764 bright galaxies at each redshift,
  which results in the samples with the same number density.

\section{RESULTS}\label{results}


Fig. \ref{z0} shows the dependence of physical parameters of galaxies
 as a function of local density ($n_5$) at $z=0.0$.
There could be many ways to determine the environment of galaxies \citep{muldrew12}, 
 but we take a simple approach as often used in 
 the analysis of observational data (e.g. \citealt{bal06,hwa10lirg}).
The local density here is the galaxy number density 
  measured in a volume of a sphere 
  defined by the distance to the 5th nearest neighbour galaxy.
The physical scale for the local environment probed by this measure
  varies with the galaxy number density of the sample, 
  but for our samples the physical scale corresponds to $0.1-10.6$ cMpc. 
The top left panel shows that 
  the SFR of individual galaxies, on average, decreases
  with local density as expected.
The stellar mass in the middle left panel increases with density.
In the result, the specific SFR (SFR per stellar mass, bottom left) decreases 
  with local density,
  consistent with observational results in the local universe (e.g. \citealt{bal04}).

\begin{figure*}
	\centering
	\includegraphics[scale=0.85]{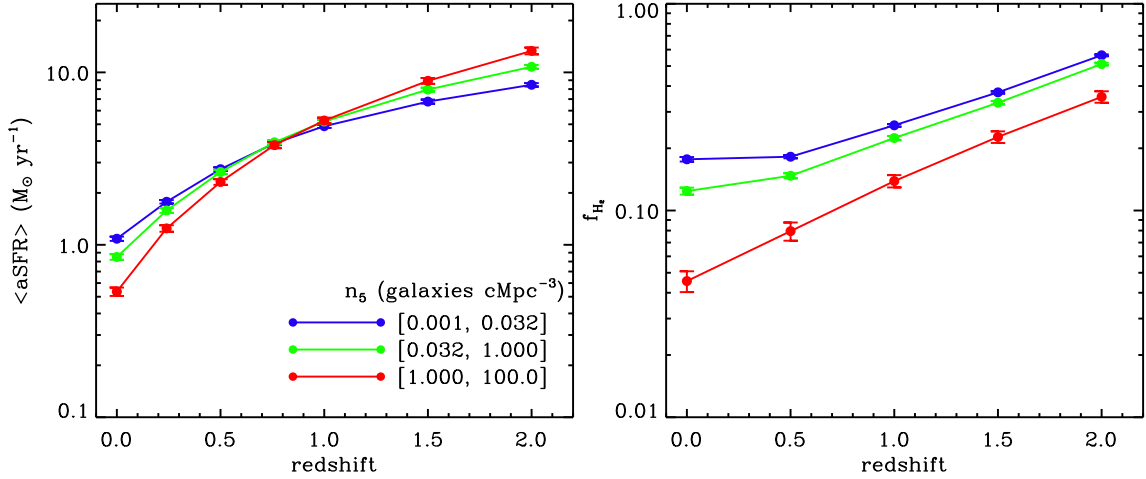}
	\caption{(Left) Evolution of average SFRs as a function of redshift
		for different local density bins.
		(Right) Similar to left panel, but molecular hydrogen mass fraction.}
	\label{zsfr}
\end{figure*}

We also compute the spatially averaged SFRs, <aSFR> (top right panel),
  which was used for the finding of the reversal of
  SFR-density relation in \citet{elb07}.
This is the mean SFR of the galaxies within a sphere of 1 cMpc 
  centred on a target galaxy (not volume average, but number average), 
  which is different from a simple mean SFR
  in bins of local density (e.g. red points in top left panel).
This quantity is similar to a smoothed SFR in a given position.
Thus, it shows a correlation between the star formation activity
  and the local density better than the SFR of individual galaxies
  that can sometimes be very noisy.
The top right panel of Fig. \ref{z0} shows a trend for such an average SFR;
  it decreases with local density,
  similar to observational data in the local universe 
  (e.g. Fig. 8 in \citealt{elb07} and Fig. 17 in \citealt{hwa10lirg}).
The mass fraction of cold gas (i.e. $f_{\rm H_2} = M_{\rm H_2}/M_{\rm star}$)
  in the middle right panel decreases with local density,
  consistent with observations (e.g. \citealt{gro16}).

Fig. \ref{z1} is similar to Fig. \ref{z0}, but for the data at $z=1.0$.
The galaxy number density and the methods calculating all the parameters
  are consistent with those at $z=0.0$.
Interestingly,
 the SFRs of individual galaxies do not decrease with local density (top left panel),
 which is different from those at $z=0.0$.
However, 
  the stellar masses and the specific SFRs, 
  respectively, increase and decrease with local density 
  (middle and bottom left panels), which is similar to those at $z=0.0$.
The change of SFRs with local density is more prominent in the plot of 
  average SFRs (top right panel),
  which shows an increase of SFR with local density (i.e. reversal of SFR-density relation)
  as seen in the observational data.
The trends for higher redshifts (i.e. $z>1.0$) are similar to the case at $z=1.0$
  (see Appendix \ref{appenA} for more plots), 
  but the increase of SFRs with local density is more pronounced.
Interestingly, the specific SFRs do not increase with local density even at high redshifts,
  which could suggest that the reversal of SFR-density relation is solely driven
  by the change of stellar mass with local density;
  we discuss this effect in detail in Section \ref{cause}.
  
  
To summarize the change of SFR-density relation with redshift,
  we show, in Fig. \ref{asfr}, the SFRs (top left) and the average SFRs (top right) of galaxies 
  as a function of local density for all the redshifts we analyse at $z=0.0-2.0$.
For comparison, we also show the change of SFRs of the SDSS galaxies 
  with local density as a grey line,
  which is adopted from Fig. 17 of \citet{hwa10lirg}.
It should be noted that the methods to compute the spatially averaged SFRs
  and the local density are not perfectly the same
  between simulation and observational data (e.g. \citealt{elb07,hwa10lirg}).
The main difference between the two is 
  the volume used for computing the average SFR and the local density.
For example, in simulations <aSFR> is the mean SFR of the galaxies within a sphere of 1 cMpc 
  centred on a target galaxy, but in observations it is the mean SFR 
  of the galaxies that have velocities relative to a target galaxy less than 1500 km s$^{-1}$and
  the projected distance less than 1 cMpc.
This difference is simply because we use the snapshot data of the TNG simulations, 
  which do not provide the radial velocities of galaxies from an observer.
This requires lightcone data, which are not yet available from the TNG simulations.
Therefore, the grey line for the SDSS data should not be directly compared with
  the lines derived from the TNG data, but should be only used as a guide line.

The top panels of Fig. \ref{asfr} show that 
  both individual and average SFRs decrease with local density at low redshifts (reddish lines).
Then, the dependence gradually changes with redshift,
  and finally the SFR increases with local density at high redshifts (i.e. $z\gtrsim1$):
  reversal of SFR-density relation. 
The bottom left panel shows the molecular gas fraction of individual galaxies
  as a function of local density for different redshifts. 
At all redshifts, the gas fraction decreases with local density.
The gas fraction systematically increases with redshift,
  consistent with observations (e.g. \citealt{sco17});
  this increase is known to be responsible for the systematic increase of specific SFRs 
  of galaxies with redshift 
  (e.g. Fig. 18 in \citealt{elb11}; \citealt{sar14}).
However, the increase of gas fraction with redshift
  is slightly faster in high-density regions than in low-density regions;
  the gas fraction at the lowest-density bin increases $\sim$3 times from $z=0.0$ to $z=2.0$, 
  but the fraction at the highest-density bin increases $\sim$10 times.

\begin{figure*}
	\centering
	\includegraphics[scale=0.8]{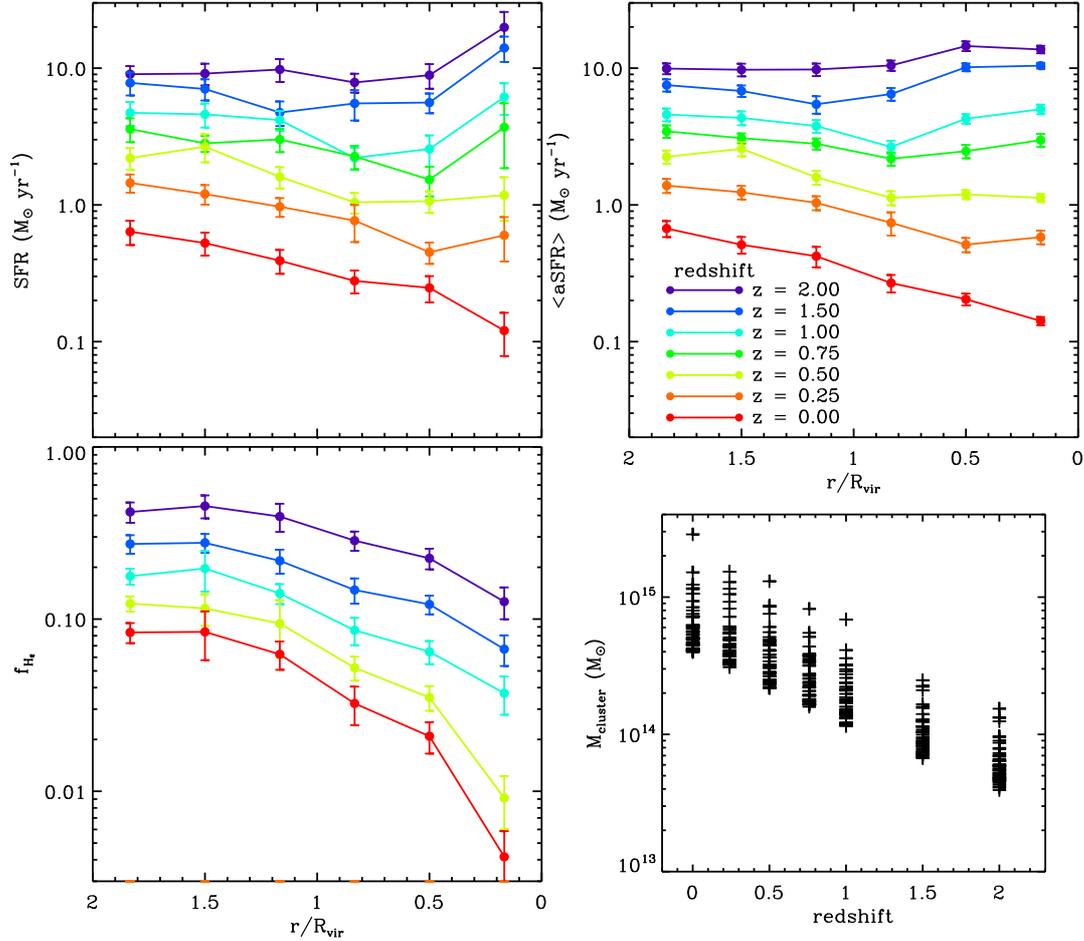}
	\caption{Similar to Fig. \ref{asfr}, but for the galaxies centred on galaxy clusters at several redshifts.
		The clustercentric radius is normalized by a virial radius of each cluster.		
		(bottom right) Cluster mass as a function of redshift.}
	\label{asfrcl}
\end{figure*}

To show the change of environmental dependence of SFRs with redshift in a different way,
  we plot the average SFRs of galaxies as a function of redshift 
  for three local density bins in the left panel of Fig. \ref{zsfr}.
The average SFR increases with redshift for all density bins,
  consistent with the expectation from an increase of gas fraction with redshift 
  (e.g. \citealt{sco17}).
However, the increase is steeper for high-density regions.
Thus, at low redshifts, the average SFR is lower in high-density regions than in low-density regions,
  but at high redshifts, the SFR is higher in high-density regions than in low-density regions.
In the result, there is a crossing of SFRs between different density bins
  around $z\sim0.7$; this corresponds to the redshift when the SFR-density relation is almost flat
  (see top panels of Fig. \ref{asfr}).
At $z\gtrsim0.7$, the average SFRs of galaxies in high-density regions are always higher
  than those of galaxies in low-density regions,
  which is the reversal of the SFR-density relation.
The right panel shows the change of gas fraction with redshift.
The gas fraction is always higher in lower density regions than in higher density regions
  regardless of redshift at $z=0.0-2.0$,
  but the difference between low and high-density regions decreases with redshift.

\section{DISCUSSION}\label{discuss}

We use the  IllustrisTNG cosmological hydrodynamical simulation 
  to study how the SFR-density relation evolves over cosmic time at $z=0.0-2.0$.
We could successfully reproduce the reversal of SFR-density 
  relation observed for high-redshift galaxies at $z\sim1.0$ \citep{elb07}.

We study the evolution of SFR-density relation in two ways.
One is to examine the SFR-density relation at different redshifts.
We therefore use several snapshot data provided by the TNG simulation, and
  systematically look into the change of the relation with redshift (e.g. Fig. \ref{asfr}).
The average SFR of galaxies decreases with local density at low redshift,
  but becomes almost flat at $z\sim0.7$.
The SFR increases with density since then, and the dependence becomes
  stronger with redshift.
The other way is to examine the change of average SFR
  with redshift by fixing the local density (e.g. Fig. \ref{zsfr}).
The average SFR of galaxies increases with redshift for any density bins, 
  but the increase is much faster in high-density regions.
Then, the crossing of SFRs between low- and high-density regions
  happens at $z\sim0.7$ when the SFR-density relation is almost flat.
At $z\gtrsim0.7$, the average SFRs of high-density regions
  are always higher than those of low-density regions, and 
  the difference between high- and low-density regions becomes larger with redshift.

\begin{figure*}
	\centering
	\includegraphics[scale=0.8]{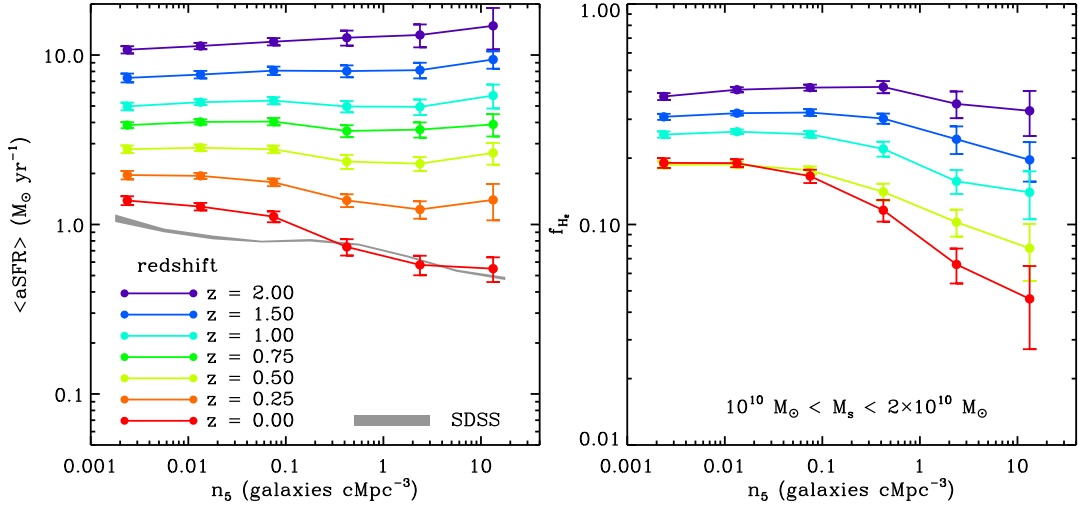}
	\caption{Similar to Fig. \ref{asfr}, but based on the galaxy sample in a very narrow
		mass range  (i.e. 10$^{10}<M_s/M_\odot<2\times10^{10}$).}
	\label{asfrmass}
\end{figure*}

\subsection{Evolution of SFR-density Relation in the Galaxy Cluster Environment}\label{cluster}
  
The crossing of SFRs between different environments 
  is also expected from a comparison of SFRs (or infrared luminosities) 
  between field and cluster galaxies
  even though the crossing redshift can vary depending on the galaxy sample
  (e.g. Fig. 5 of \citealt{alb14}).
In this regard, to better examine the environmental dependence of SFRs in very high-density regions
  that are not often resolved with typical local density indicators (e.g. $n_5$ in this study),
  we show the dependence of SFRs and the mass fraction of molecular hydrogen
  on the distance to a galaxy cluster in Fig. \ref{asfrcl}.
The top left panel shows the SFRs of individual galaxies
  as a function of distance to a galaxy cluster.
The galaxy sample at each redshift is the same as the one used in the previous section.
To calculate the distance to a galaxy cluster for each galaxy in the sample,
  we first identify 50 most massive galaxy clusters at each redshift 
  from the halo catalogues in the TNG300-1
  (see the bottom right panel for the mass distribution of the cluster sample).
We then plot the SFRs of the galaxies close to those massive clusters 
  as a function of distance to a galaxy cluster (normalized by virial radii of clusters).

As expected, the SFRs of the galaxies at low redshifts
  decrease with decreasing clustercentric radius,
  consistent with observational results (e.g. \citealt{bg06,ph09, von10}).
Similar to the dependence on local density,   
  the dependence of SFR changes with redshifts,
  but the change is not as dramatic as the case based on the local density.
Even at $z=2.0$, 
  the SFR remains almost flat at $r/R_{\rm vir}\gtrsim0.5$, and increases only
  at $r/R_{\rm vir}<0.5$.
The average SFRs in the top right panel shows a similar trend.
Interestingly, the SFRs in high-density regions (i.e. $r/R_{\rm vir}<0.5$)
  are systematically higher than those in low-density regions
  only at $z\geq1.5$.
This redshift is higher than the redshift that shows
  the reversal of SFR-density relation based on local density (i.e. $z\geq1.0$);
  this difference will be more discussed later in this section.
The mass fraction of molecular hydrogen in the bottom left panel 
  also decreases with decreasing clustercentric radius at all redshifts.
The increase of molecular hydrogen mass fraction with redshift
  is more prominent in high-density regions, 
  consistent with the results based on local density.

\subsection{What Causes the Reversal of the SFR-density Relation?}\label{cause}

Galaxy properties are strongly affected by both internal (e.g. stellar mass) and 
  external factors (e.g. environment) \citep{peng10},
  which suggests that
  the SFR-density relation may not be simply the outcome of environmental effects. 
To examine whether the SFR-density relation is mainly driven by the difference in
  stellar mass of galaxies at different density bins, we plot in Fig. \ref{asfrmass}
  the average SFRs as a function of local density for different redshifts
  by narrowing the mass range of the sample galaxies 
  (i.e. 10$^{10}<M_s/M_\odot<2\times10^{10}$; 
  blue horizontal bands in the middle left panels of Figs. \ref{z0} and \ref{z1}).
The left panel shows that
  the SFR-density relation still changes with redshift
  despite the small mass range, indicating a clear role of galaxy environment in the relation.

As the mass fraction of molecular hydrogen in high-density regions 
  decreases with decreasing redshift faster than in low-density regions,
  one can expect that the environment-dependent evolution of gas fraction
  is responsible for the reversal of the SFR-density relation.
According to the bottom left panel of Figure \ref{asfr},
  the gas fraction at the lowest-density bin increases $\sim$3 times from $z = 0.0$ to $z = 2.0$, 
  but the fraction at the highest-density bin increases $\sim$10 times. 
This can imply an increase of SFR by a factor of ($10/3$)$^{1.4}\approx5$
  if we assume a simple Schmidt-Kennicutt law with a slope of 1.4  
  for the conversion between gas and star formation rate densities \citep{sch59,ken98law}.
When we compare the change of SFR between lowest- and highest-density bins 
  in the top panels of Figure \ref{asfr}, 
  the ratio of increase in SFR between the two density bins is $\sim5$.
This ratio is consistent with the one for the increase in gas fraction, which 
  can suggest that the change of gas fraction is responsible for 
  the evolution of environment-dependent SFR.
However, this does not fully explain the trend.
It should be noted that 
  the mass fraction of molecular hydrogen is always lower in high-density regions 
  than in low-density regions even at high redshifts
  when the average SFR in high-density regions is higher than 
  in low-density regions.
In addition, even when we fix the stellar mass, 
  the average SFR still increases with density at $z\geq1.5$, but 
  the gas fraction changes little with local density (right panel of Fig. \ref{asfrmass};
  see also \citealt{dar18} for similar observational results).   
This suggests that there would be another factor in addition to the gas fraction, 
  which affects the star formation activity of galaxies: e.g. environment.
In other words, there should be a mechanism,
  which can make the overall star formation rates in high-density regions
  higher than those in low-density regions at high redshifts 
  (i.e. reversal of the SFR-density relation) 
  even though the gas fraction in high-density regions is not higher than
  that in low-density regions.
 
As \citet{hwa10lirg} and \citet{tc14} discussed,
   an important factor behind the evolution of the SFR-density relation
   is how the fraction of massive, star-forming galaxies changes with local density 
   through cosmic time.
This could be related to the observational finding
  that the star-formation main sequence shows a bending at high masses 
  and the bending evolves with redshift \citep{whi15,sch15ms}. 
Here the bending means that the SFRs of massive star-forming galaxies
  are lower than the extrapolations from the best-fit relation for less massive galaxies.
The bending of the main sequence is more pronounced at lower redshifts,
  which means that the difference between 
  the measured SFRs of more massive galaxies
  and the expected ones from less massive galaxies 
  increases with decreasing redshift.
As high-density regions tend to host more massive galaxies
  and the bending of the main sequence is more pronounced at lower redshifts,
  the decrease of average SFRs with redshift in high-density regions 
  where more massive galaxies exist can be accelerated than 
  that in low-density regions.

There are several explanations for the bending of 
  the star-formation main sequence,
  which include a growth of bulge component as a consequence of 
  the increased merger activity in groups and 
  an environment-driven gas removal/consumption process \citep{pop19}.  
In the hierarchical picture of galaxy formation, 
  these phenomena could be understood as the results of continuous 
  interactions and mergers with other galaxies.
In more detail, at high redshifts, galaxies overall have large amounts of cold gas,
  which can be easily converted into stars.
Then, the galaxies in high-density regions consume their gas 
	faster than those in low-density regions 
	through galaxy-galaxy interactions and merging  \citep{hwa11inter,hwa12shels,pan18}
	that are generally expected to be more frequent than in low-density regions.
Therefore, in high-density regions at high redshifts, 
  there could be many massive, star-forming galaxies,
  forming larger structures such as galaxy groups/clusters;
  indeed, massive, star-forming galaxies are strongly clustered at high redshifts
  (e.g., \citealt{far06,gil07}).
In the end, at low redshifts, those galaxies 
  become gas poor and quiescent, being dominant in high-density regions.

When we focus on very high-density regions such as galaxy clusters,
  various cluster-driven quenching mechanisms 
  related to the hot gas of clusters  \citep{bg06} are
  expected to play a role in addition to 
  the processes mentioned above.
Therefore, the decrease of SFRs in cluster regions starts earlier than
   any other environments.
This is  what the top panels of Fig. \ref{asfrcl} shows;
  the SFRs in the central regions of clusters (i.e. $r/R_{\rm vir}<0.5$)
  are systematically higher than those in the other regions only at $z\geq1.5$,
  which is a higher redshift than the redshift that shows
  the reversal of the SFR-density relation based on local density (i.e. $z\geq1.0$).
This can explain why some studies do not find obvious increase
  of SFRs with decreasing clustercentric radius at high redshifts ($z\gtrsim1$; e.g.
  \citealt{bro13,alb16}) and/or do find quenching of star formation activity 
  in the central regions of clusters at $z\sim1.5$ \citep{str19}
  despite the expectations from the increase of SFRs with local density.

The environmental dependent evolution of star formation activity of galaxies 
  is consistent with the observational results for other galaxy properties 
  (e.g., \citealt{cap07,hp09}).
For example, \citet{hp09} showed that
  the early-type galaxy fraction changes much faster in high-density regions
  than in low-density regions, which results in a weaker correlation 
  between galaxy morphology and local density at high redshifts than at low redshifts.
Again, this could be the results of frequent galaxy-galaxy interactions 
   in higher density regions at high redshifts that
   accelerate the morphological transformation of late-type galaxies into early-type galaxies
   along with the quenching of star formation activity.
    
\section{SUMMARY}\label{sum}

We use the IllustrisTNG cosmological hydrodynamical simulation
  to study the evolution of SFR-density relation at $z=2.0-0.0$, and
  successfully reproduce the observed reversal of SFR-density relation at high redshifts.
Our primary results are:

\begin{enumerate}
\item The SFR of galaxies decreases with local density at $z=0.0$, 
  but its dependence on local density becomes weaker with redshift.
At $z\gtrsim1.0$, the SFR of galaxies increases with local density
  (i.e. reversal of SFR-density relation),  
  and its dependence becomes stronger with redshift.
This change of SFR-density relation with redshift still remains 
  even when fixing the stellar masses of galaxies.

\item The dependence of SFR on the distance to a galaxy cluster shows a trend
  similar to the one based on local density.
However, the change of the dependence with redshift 
  is not as dramatic as the one based on local density.
This difference could be because
  in clusters there are cluster-driven quenching mechanisms
  in addition to the processes related to
  frequent galaxy-galaxy interactions and merging.
  
\item The gas fraction of molecular hydrogen always 
  decreases with local density regardless of redshift  at $z=0.0-2.0$
  even though the dependence becomes weaker when we fix the stellar mass.

\end{enumerate}
	
Our results are consistent with the idea that
  massive, star-forming galaxies are strongly clustered at high redshifts,
  forming larger structures.
They then consume their gas faster than field galaxies 
  through frequent interactions with other galaxies,
  ending up being quiescent in the local universe.
We plan to study in detail the connection between galaxy interactions
  and the dramatic change of environmental dependence of galaxy properties
  using the cosmological hydrodynamical simulations including the TNG.

\section*{Acknowledgements}

We thank the referee for constructive comments.
We thank the Korea Institute for Advanced Study for providing computing resources (KIAS Center for Advanced Computation) for this work.
We thank the IllustrisTNG collaboration for making their simulation data publicly available.





\bibliographystyle{mnras} 
\bibliography{/Users/hhwang/GoogleDrive/0_Papers_Working/ref_hshwang} 

\begin{thebibliography}{}
\makeatletter
\relax
\def\mn@urlcharsother{\let\do\@makeother \do\$\do\&\do\#\do\^\do\_\do\%\do\~}
\def\mn@doi{\begingroup\mn@urlcharsother \@ifnextchar [ {\mn@doi@}
  {\mn@doi@[]}}
\def\mn@doi@[#1]#2{\def\@tempa{#1}\ifx\@tempa\@empty \href
  {http://dx.doi.org/#2} {doi:#2}\else \href {http://dx.doi.org/#2} {#1}\fi
  \endgroup}
\def\mn@eprint#1#2{\mn@eprint@#1:#2::\@nil}
\def\mn@eprint@arXiv#1{\href {http://arxiv.org/abs/#1} {{\tt arXiv:#1}}}
\def\mn@eprint@dblp#1{\href {http://dblp.uni-trier.de/rec/bibtex/#1.xml}
  {dblp:#1}}
\def\mn@eprint@#1:#2:#3:#4\@nil{\def\@tempa {#1}\def\@tempb {#2}\def\@tempc
  {#3}\ifx \@tempc \@empty \let \@tempc \@tempb \let \@tempb \@tempa \fi \ifx
  \@tempb \@empty \def\@tempb {arXiv}\fi \@ifundefined
  {mn@eprint@\@tempb}{\@tempb:\@tempc}{\expandafter \expandafter \csname
  mn@eprint@\@tempb\endcsname \expandafter{\@tempc}}}

\bibitem[\protect\citeauthoryear{{Alberts} et~al.,}{{Alberts}
  et~al.}{2014}]{alb14}
{Alberts} S.,  et~al., 2014, \mn@doi [\mnras] {10.1093/mnras/stt1897}, \href
  {https://ui.adsabs.harvard.edu/\#abs/2014MNRAS.437..437A} {437, 437}

\bibitem[\protect\citeauthoryear{{Alberts} et~al.,}{{Alberts}
  et~al.}{2016}]{alb16}
{Alberts} S.,  et~al., 2016, \mn@doi [\apj] {10.3847/0004-637X/825/1/72}, \href
  {https://ui.adsabs.harvard.edu/\#abs/2016ApJ...825...72A} {825, 72}

\bibitem[\protect\citeauthoryear{{Baldry}, {Balogh}, {Bower}, {Glazebrook},
  {Nichol}, {Bamford}  \& {Budavari}}{{Baldry} et~al.}{2006}]{bal06}
{Baldry} I.~K.,  {Balogh} M.~L.,  {Bower} R.~G.,  {Glazebrook} K.,  {Nichol}
  R.~C.,  {Bamford} S.~P.,   {Budavari} T.,  2006, \mn@doi [\mnras]
  {10.1111/j.1365-2966.2006.11081.x}, \href
  {https://ui.adsabs.harvard.edu/abs/2006MNRAS.373..469B} {373, 469}

\bibitem[\protect\citeauthoryear{{Balogh}, {Schade}, {Morris}, {Yee},
  {Carlberg}  \& {Ellingson}}{{Balogh} et~al.}{1998}]{bal98}
{Balogh} M.~L.,  {Schade} D.,  {Morris} S.~L.,  {Yee} H.~K.~C.,  {Carlberg}
  R.~G.,   {Ellingson} E.,  1998, \mn@doi [\apj] {10.1086/311576}, \href
  {https://ui.adsabs.harvard.edu/abs/1998ApJ...504L..75B} {504, L75}

\bibitem[\protect\citeauthoryear{{Balogh} et~al.,}{{Balogh}
  et~al.}{2004}]{bal04}
{Balogh} M.,  et~al., 2004, \mn@doi [\mnras]
  {10.1111/j.1365-2966.2004.07453.x}, \href
  {http://adsabs.harvard.edu/abs/2004MNRAS.348.1355B} {348, 1355}

\bibitem[\protect\citeauthoryear{{Blanton} \& {Moustakas}}{{Blanton} \&
  {Moustakas}}{2009}]{bm09}
{Blanton} M.~R.,  {Moustakas} J.,  2009, \mn@doi [\araa]
  {10.1146/annurev-astro-082708-101734}, \href
  {http://adsabs.harvard.edu/abs/2009ARA%26A..47..159B} {47, 159}

\bibitem[\protect\citeauthoryear{{Blanton}, {Eisenstein}, {Hogg}, {Schlegel}
  \& {Brinkmann}}{{Blanton} et~al.}{2005}]{bla05cdr}
{Blanton} M.~R.,  {Eisenstein} D.,  {Hogg} D.~W.,  {Schlegel} D.~J.,
  {Brinkmann} J.,  2005, \mn@doi [\apj] {10.1086/422897}, \href
  {http://adsabs.harvard.edu/abs/2005ApJ...629..143B} {629, 143}

\bibitem[\protect\citeauthoryear{{Boselli} \& {Gavazzi}}{{Boselli} \&
  {Gavazzi}}{2006}]{bg06}
{Boselli} A.,  {Gavazzi} G.,  2006, \mn@doi [\pasp] {10.1086/500691}, \href
  {http://adsabs.harvard.edu/abs/2006PASP..118..517B} {118, 517}

\bibitem[\protect\citeauthoryear{{Brodwin} et~al.,}{{Brodwin}
  et~al.}{2013}]{bro13}
{Brodwin} M.,  et~al., 2013, \mn@doi [\apj] {10.1088/0004-637X/779/2/138},
  \href {https://ui.adsabs.harvard.edu/\#abs/2013ApJ...779..138B} {779, 138}

\bibitem[\protect\citeauthoryear{{Capak}, {Abraham}, {Ellis}, {Mobasher},
  {Scoville}, {Sheth}  \& {Koekemoer}}{{Capak} et~al.}{2007}]{cap07}
{Capak} P.,  {Abraham} R.~G.,  {Ellis} R.~S.,  {Mobasher} B.,  {Scoville} N.,
  {Sheth} K.,   {Koekemoer} A.,  2007, \mn@doi [\apjs] {10.1086/518424}, \href
  {http://adsabs.harvard.edu/abs/2007ApJS..172..284C} {172, 284}

\bibitem[\protect\citeauthoryear{{Cooper} et~al.,}{{Cooper}
  et~al.}{2007}]{coo07}
{Cooper} M.~C.,  et~al., 2007, \mn@doi [\mnras]
  {10.1111/j.1365-2966.2007.11534.x}, \href
  {http://adsabs.harvard.edu/abs/2007MNRAS.376.1445C} {376, 1445}

\bibitem[\protect\citeauthoryear{{Cooper} et~al.,}{{Cooper}
  et~al.}{2008}]{coo08}
{Cooper} M.~C.,  et~al., 2008, \mn@doi [\mnras]
  {10.1111/j.1365-2966.2007.12613.x}, \href
  {http://adsabs.harvard.edu/abs/2008MNRAS.383.1058C} {383, 1058}

\bibitem[\protect\citeauthoryear{{Cucciati} et~al.,}{{Cucciati}
  et~al.}{2006}]{cuc06}
{Cucciati} O.,  et~al., 2006, \mn@doi [\aap] {10.1051/0004-6361:20065161},
  \href {https://ui.adsabs.harvard.edu/abs/2006A&A...458...39C} {458, 39}

\bibitem[\protect\citeauthoryear{{Darvish}, {Scoville}, {Martin}, {Mobasher},
  {Diaz-Santos}  \& {Shen}}{{Darvish} et~al.}{2018}]{dar18}
{Darvish} B.,  {Scoville} N.~Z.,  {Martin} C.,  {Mobasher} B.,  {Diaz-Santos}
  T.,   {Shen} L.,  2018, \mn@doi [\apj] {10.3847/1538-4357/aac836}, \href
  {https://ui.adsabs.harvard.edu/abs/2018ApJ...860..111D} {860, 111}

\bibitem[\protect\citeauthoryear{{Diemer} et~al.,}{{Diemer}
  et~al.}{2018}]{die18}
{Diemer} B.,  et~al., 2018, \mn@doi [The Astrophysical Journal Supplement
  Series] {10.3847/1538-4365/aae387}, \href
  {https://ui.adsabs.harvard.edu/abs/2018ApJS..238...33D} {238, 33}

\bibitem[\protect\citeauthoryear{{Diemer} et~al.,}{{Diemer}
  et~al.}{2019}]{die19}
{Diemer} B.,  et~al., 2019, \mn@doi [\mnras] {10.1093/mnras/stz1323}, \href
  {https://ui.adsabs.harvard.edu/abs/2019MNRAS.487.1529D} {487, 1529}

\bibitem[\protect\citeauthoryear{{Dressler}}{{Dressler}}{1980}]{dre80}
{Dressler} A.,  1980, \mn@doi [\apj] {10.1086/157753}, \href
  {http://adsabs.harvard.edu/abs/1980ApJ...236..351D} {236, 351}

\bibitem[\protect\citeauthoryear{{Dubois} et~al.,}{{Dubois}
  et~al.}{2014}]{dub14}
{Dubois} Y.,  et~al., 2014, \mn@doi [\mnras] {10.1093/mnras/stu1227}, \href
  {http://adsabs.harvard.edu/abs/2014MNRAS.444.1453D} {444, 1453}

\bibitem[\protect\citeauthoryear{{Elbaz} et~al.,}{{Elbaz} et~al.}{2007}]{elb07}
{Elbaz} D.,  et~al., 2007, \mn@doi [\aap] {10.1051/0004-6361:20077525}, \href
  {http://adsabs.harvard.edu/abs/2007A%26A...468...33E} {468, 33}

\bibitem[\protect\citeauthoryear{{Elbaz} et~al.,}{{Elbaz} et~al.}{2011}]{elb11}
{Elbaz} D.,  et~al., 2011, \mn@doi [\aap] {10.1051/0004-6361/201117239}, \href
  {http://adsabs.harvard.edu/abs/2011A%26A...533A.119E} {533, 119}

\bibitem[\protect\citeauthoryear{{Farrah} et~al.,}{{Farrah}
  et~al.}{2006}]{far06}
{Farrah} D.,  et~al., 2006, \mn@doi [\apjl] {10.1086/503769}, \href
  {http://adsabs.harvard.edu/abs/2006ApJ...641L..17F} {641, L17}

\bibitem[\protect\citeauthoryear{{Giavalisco} et~al.,}{{Giavalisco}
  et~al.}{2004}]{gia04}
{Giavalisco} M.,  et~al., 2004, \mn@doi [\apjl] {10.1086/379232}, \href
  {http://adsabs.harvard.edu/abs/2004ApJ...600L..93G} {600, L93}

\bibitem[\protect\citeauthoryear{{Gilli} et~al.,}{{Gilli} et~al.}{2007}]{gil07}
{Gilli} R.,  et~al., 2007, \mn@doi [\aap] {10.1051/0004-6361:20077506}, \href
  {http://adsabs.harvard.edu/abs/2007A%26A...475...83G} {475, 83}

\bibitem[\protect\citeauthoryear{{Gnedin} \& {Draine}}{{Gnedin} \&
  {Draine}}{2014}]{gd14}
{Gnedin} N.~Y.,  {Draine} B.~T.,  2014, \mn@doi [\apj]
  {10.1088/0004-637X/795/1/37}, \href
  {https://ui.adsabs.harvard.edu/abs/2014ApJ...795...37G} {795, 37}

\bibitem[\protect\citeauthoryear{{G{\'o}mez} et~al.,}{{G{\'o}mez}
  et~al.}{2003}]{gom03}
{G{\'o}mez} P.~L.,  et~al., 2003, \mn@doi [\apj] {10.1086/345593}, \href
  {http://adsabs.harvard.edu/abs/2003ApJ...584..210G} {584, 210}

\bibitem[\protect\citeauthoryear{{Grossi} et~al.,}{{Grossi}
  et~al.}{2016}]{gro16}
{Grossi} M.,  et~al., 2016, \mn@doi [\aap] {10.1051/0004-6361/201628123}, \href
  {https://ui.adsabs.harvard.edu/abs/2016A&A...590A..27G} {590, A27}

\bibitem[\protect\citeauthoryear{{Grossi}, {Fernandes}, {Sobral}, {Afonso},
  {Telles}, {Bizzocchi}, {Paulino-Afonso}  \& {Matute}}{{Grossi}
  et~al.}{2018}]{gro18}
{Grossi} M.,  {Fernandes} C. A.~C.,  {Sobral} D.,  {Afonso} J.,  {Telles} E.,
  {Bizzocchi} L.,  {Paulino-Afonso} A.,   {Matute} I.,  2018, \mn@doi [\mnras]
  {10.1093/mnras/stx3165}, \href
  {https://ui.adsabs.harvard.edu/abs/2018MNRAS.475..735G} {475, 735}

\bibitem[\protect\citeauthoryear{{Hogg} et~al.,}{{Hogg} et~al.}{2003}]{hogg03}
{Hogg} D.~W.,  et~al., 2003, \mn@doi [\apj] {10.1086/374238}, \href
  {https://ui.adsabs.harvard.edu/abs/2003ApJ...585L...5H} {585, L5}

\bibitem[\protect\citeauthoryear{{Hwang} \& {Park}}{{Hwang} \&
  {Park}}{2009}]{hp09}
{Hwang} H.~S.,  {Park} C.,  2009, \mn@doi [\apj] {10.1088/0004-637X/700/1/791},
  \href {http://adsabs.harvard.edu/abs/2009ApJ...700..791H} {700, 791}

\bibitem[\protect\citeauthoryear{{Hwang}, {Elbaz}, {Lee}, {Jeong}, {Park},
  {Lee}  \& {Lee}}{{Hwang} et~al.}{2010}]{hwa10lirg}
{Hwang} H.~S.,  {Elbaz} D.,  {Lee} J.~C.,  {Jeong} W.,  {Park} C.,  {Lee}
  M.~G.,   {Lee} H.~M.,  2010, \mn@doi [\aap] {10.1051/0004-6361/201014807},
  \href {http://adsabs.harvard.edu/abs/2010A%26A...522A..33H} {522, 33}

\bibitem[\protect\citeauthoryear{{Hwang} et~al.,}{{Hwang}
  et~al.}{2011}]{hwa11inter}
{Hwang} H.~S.,  et~al., 2011, \mn@doi [\aap] {10.1051/0004-6361/201117476},
  \href {http://adsabs.harvard.edu/abs/2011A%26A...535A..60H} {535, 60}

\bibitem[\protect\citeauthoryear{{Hwang}, {Geller}, {Kurtz}, {Dell'Antonio}  \&
  {Fabricant}}{{Hwang} et~al.}{2012}]{hwa12shels}
{Hwang} H.~S.,  {Geller} M.~J.,  {Kurtz} M.~J.,  {Dell'Antonio} I.~P.,
  {Fabricant} D.~G.,  2012, \mn@doi [\apj] {10.1088/0004-637X/758/1/25}, \href
  {http://adsabs.harvard.edu/abs/2012ApJ...758...25H} {758, 25}

\bibitem[\protect\citeauthoryear{{Kennicutt}}{{Kennicutt}}{1998}]{ken98law}
{Kennicutt} R.~C. J.,  1998, \mn@doi [\apj] {10.1086/305588}, \href
  {http://adsabs.harvard.edu/abs/1998ApJ...498..541K} {498, 541}

\bibitem[\protect\citeauthoryear{{Khandai}, {Di Matteo}, {Croft}, {Wilkins},
  {Feng}, {Tucker}, {DeGraf}  \& {Liu}}{{Khandai} et~al.}{2015}]{kha15}
{Khandai} N.,  {Di Matteo} T.,  {Croft} R.,  {Wilkins} S.,  {Feng} Y.,
  {Tucker} E.,  {DeGraf} C.,   {Liu} M.-S.,  2015, \mn@doi [\mnras]
  {10.1093/mnras/stv627}, \href
  {http://adsabs.harvard.edu/abs/2015MNRAS.450.1349K} {450, 1349}

\bibitem[\protect\citeauthoryear{{Lewis} et~al.,}{{Lewis} et~al.}{2002}]{lew02}
{Lewis} I.,  et~al., 2002, \mn@doi [\mnras] {10.1046/j.1365-8711.2002.05558.x},
  \href {http://adsabs.harvard.edu/abs/2002MNRAS.334..673L} {334, 673}

\bibitem[\protect\citeauthoryear{{Marinacci} et~al.,}{{Marinacci}
  et~al.}{2018}]{mar18}
{Marinacci} F.,  et~al., 2018, \mn@doi [\mnras] {10.1093/mnras/sty2206}, \href
  {http://adsabs.harvard.edu/abs/2018MNRAS.480.5113M} {480, 5113}

\bibitem[\protect\citeauthoryear{{Muldrew} et~al.,}{{Muldrew}
  et~al.}{2012}]{muldrew12}
{Muldrew} S.~I.,  et~al., 2012, \mn@doi [\mnras]
  {10.1111/j.1365-2966.2011.19922.x}, \href
  {http://adsabs.harvard.edu/abs/2012MNRAS.419.2670M} {419, 2670}

\bibitem[\protect\citeauthoryear{{Naiman} et~al.,}{{Naiman}
  et~al.}{2018}]{nai18}
{Naiman} J.~P.,  et~al., 2018, \mn@doi [\mnras] {10.1093/mnras/sty618}, \href
  {http://adsabs.harvard.edu/abs/2018MNRAS.477.1206N} {477, 1206}

\bibitem[\protect\citeauthoryear{{Nelson} et~al.,}{{Nelson}
  et~al.}{2018}]{nel18}
{Nelson} D.,  et~al., 2018, \mn@doi [\mnras] {10.1093/mnras/stx3040}, \href
  {http://adsabs.harvard.edu/abs/2018MNRAS.475..624N} {475, 624}

\bibitem[\protect\citeauthoryear{{Nelson} et~al.,}{{Nelson}
  et~al.}{2019}]{nel19}
{Nelson} D.,  et~al., 2019, \mn@doi [Computational Astrophysics and Cosmology]
  {10.1186/s40668-019-0028-x}, \href
  {https://ui.adsabs.harvard.edu/abs/2019ComAC...6....2N} {6, 2}

\bibitem[\protect\citeauthoryear{{Pan} et~al.,}{{Pan} et~al.}{2018}]{pan18}
{Pan} H.-A.,  et~al., 2018, \mn@doi [\apj] {10.3847/1538-4357/aaeb92}, \href
  {https://ui.adsabs.harvard.edu/abs/2018ApJ...868..132P} {868, 132}

\bibitem[\protect\citeauthoryear{{Park} \& {Hwang}}{{Park} \&
  {Hwang}}{2009}]{ph09}
{Park} C.,  {Hwang} H.~S.,  2009, \mn@doi [\apj]
  {10.1088/0004-637X/699/2/1595}, \href
  {http://adsabs.harvard.edu/abs/2009ApJ...699.1595P} {699, 1595}

\bibitem[\protect\citeauthoryear{{Pelliccia} et~al.,}{{Pelliccia}
  et~al.}{2019}]{pel19}
{Pelliccia} D.,  et~al., 2019, \mn@doi [\mnras] {10.1093/mnras/sty2876}, \href
  {https://ui.adsabs.harvard.edu/abs/2019MNRAS.482.3514P} {482, 3514}

\bibitem[\protect\citeauthoryear{{Peng} et~al.,}{{Peng} et~al.}{2010}]{peng10}
{Peng} Y.-j.,  et~al., 2010, \mn@doi [\apj] {10.1088/0004-637X/721/1/193},
  \href {https://ui.adsabs.harvard.edu/abs/2010ApJ...721..193P} {721, 193}

\bibitem[\protect\citeauthoryear{{Pillepich} et~al.,}{{Pillepich}
  et~al.}{2018a}]{pil18tng}
{Pillepich} A.,  et~al., 2018a, \mn@doi [\mnras] {10.1093/mnras/stx2656}, \href
  {http://adsabs.harvard.edu/abs/2018MNRAS.473.4077P} {473, 4077}

\bibitem[\protect\citeauthoryear{{Pillepich} et~al.,}{{Pillepich}
  et~al.}{2018b}]{pil18}
{Pillepich} A.,  et~al., 2018b, \mn@doi [\mnras] {10.1093/mnras/stx3112}, \href
  {http://adsabs.harvard.edu/abs/2018MNRAS.475..648P} {475, 648}

\bibitem[\protect\citeauthoryear{{Planck Collaboration} et~al.,}{{Planck
  Collaboration} et~al.}{2016}]{planck16}
{Planck Collaboration} et~al., 2016, \mn@doi [\aap]
  {10.1051/0004-6361/201525830}, \href
  {https://ui.adsabs.harvard.edu/\#abs/2016A&A...594A..13P} {594, A13}

\bibitem[\protect\citeauthoryear{{Popesso} et~al.,}{{Popesso}
  et~al.}{2019}]{pop19}
{Popesso} P.,  et~al., 2019, \mn@doi [\mnras] {10.1093/mnras/sty3210}, \href
  {https://ui.adsabs.harvard.edu/abs/2019MNRAS.483.3213P} {483, 3213}

\bibitem[\protect\citeauthoryear{{Postman} \& {Geller}}{{Postman} \&
  {Geller}}{1984}]{pg84}
{Postman} M.,  {Geller} M.~J.,  1984, \mn@doi [\apj] {10.1086/162078}, \href
  {https://ui.adsabs.harvard.edu/abs/1984ApJ...281...95P} {281, 95}

\bibitem[\protect\citeauthoryear{{Sargent} et~al.,}{{Sargent}
  et~al.}{2014}]{sar14}
{Sargent} M.~T.,  et~al., 2014, \mn@doi [\apj] {10.1088/0004-637X/793/1/19},
  \href {https://ui.adsabs.harvard.edu/abs/2014ApJ...793...19S} {793, 19}

\bibitem[\protect\citeauthoryear{{Schaye} et~al.,}{{Schaye}
  et~al.}{2015}]{sch15}
{Schaye} J.,  et~al., 2015, \mn@doi [\mnras] {10.1093/mnras/stu2058}, \href
  {http://adsabs.harvard.edu/abs/2015MNRAS.446..521S} {446, 521}

\bibitem[\protect\citeauthoryear{{Schmidt}}{{Schmidt}}{1959}]{sch59}
{Schmidt} M.,  1959, \mn@doi [\apj] {10.1086/146614}, \href
  {http://adsabs.harvard.edu/abs/1959ApJ...129..243S} {129, 243}

\bibitem[\protect\citeauthoryear{{Schreiber} et~al.,}{{Schreiber}
  et~al.}{2015}]{sch15ms}
{Schreiber} C.,  et~al., 2015, \mn@doi [\aap] {10.1051/0004-6361/201425017},
  \href {https://ui.adsabs.harvard.edu/abs/2015A&A...575A..74S} {575, A74}

\bibitem[\protect\citeauthoryear{{Scoville} et~al.,}{{Scoville}
  et~al.}{2013}]{sco13}
{Scoville} N.,  et~al., 2013, \mn@doi [The Astrophysical Journal Supplement
  Series] {10.1088/0067-0049/206/1/3}, \href
  {https://ui.adsabs.harvard.edu/\#abs/2013ApJS..206....3S} {206, 3}

\bibitem[\protect\citeauthoryear{{Scoville} et~al.,}{{Scoville}
  et~al.}{2017}]{sco17}
{Scoville} N.,  et~al., 2017, \mn@doi [\apj] {10.3847/1538-4357/aa61a0}, \href
  {https://ui.adsabs.harvard.edu/abs/2017ApJ...837..150S} {837, 150}

\bibitem[\protect\citeauthoryear{{Springel}}{{Springel}}{2010}]{spr10}
{Springel} V.,  2010, \mn@doi [\mnras] {10.1111/j.1365-2966.2009.15715.x},
  \href {https://ui.adsabs.harvard.edu/\#abs/2010MNRAS.401..791S} {401, 791}

\bibitem[\protect\citeauthoryear{{Springel}, {White}, {Tormen}  \&
  {Kauffmann}}{{Springel} et~al.}{2001}]{spr01}
{Springel} V.,  {White} S. D.~M.,  {Tormen} G.,   {Kauffmann} G.,  2001,
  \mn@doi [\mnras] {10.1046/j.1365-8711.2001.04912.x}, \href
  {https://ui.adsabs.harvard.edu/abs/2001MNRAS.328..726S} {328, 726}

\bibitem[\protect\citeauthoryear{{Springel} et~al.,}{{Springel}
  et~al.}{2018}]{spr18}
{Springel} V.,  et~al., 2018, \mn@doi [\mnras] {10.1093/mnras/stx3304}, \href
  {http://adsabs.harvard.edu/abs/2018MNRAS.475..676S} {475, 676}

\bibitem[\protect\citeauthoryear{{Strazzullo} et~al.,}{{Strazzullo}
  et~al.}{2019}]{str19}
{Strazzullo} V.,  et~al., 2019, \mn@doi [\aap] {10.1051/0004-6361/201833944},
  \href {https://ui.adsabs.harvard.edu/abs/2019A&A...622A.117S} {622, A117}

\bibitem[\protect\citeauthoryear{{Tonnesen} \& {Cen}}{{Tonnesen} \&
  {Cen}}{2014}]{tc14}
{Tonnesen} S.,  {Cen} R.,  2014, \mn@doi [\apj] {10.1088/0004-637X/788/2/133},
  \href {https://ui.adsabs.harvard.edu/\#abs/2014ApJ...788..133T} {788, 133}

\bibitem[\protect\citeauthoryear{{Tran} et~al.,}{{Tran} et~al.}{2010}]{tran10}
{Tran} K.-V.~H.,  et~al., 2010, \mn@doi [\apj] {10.1088/2041-8205/719/2/L126},
  \href {https://ui.adsabs.harvard.edu/abs/2010ApJ...719L.126T} {719, L126}

\bibitem[\protect\citeauthoryear{{Vogelsberger} et~al.,}{{Vogelsberger}
  et~al.}{2014}]{vog14}
{Vogelsberger} M.,  et~al., 2014, \mn@doi [\mnras] {10.1093/mnras/stu1536},
  \href {http://adsabs.harvard.edu/abs/2014MNRAS.444.1518V} {444, 1518}

\bibitem[\protect\citeauthoryear{{Whitaker} et~al.,}{{Whitaker}
  et~al.}{2015}]{whi15}
{Whitaker} K.~E.,  et~al., 2015, \mn@doi [\apj] {10.1088/2041-8205/811/1/L12},
  \href {https://ui.adsabs.harvard.edu/abs/2015ApJ...811L..12W} {811, L12}

\bibitem[\protect\citeauthoryear{{York} et~al.,}{{York} et~al.}{2000}]{york00}
{York} D.~G.,  et~al., 2000, \mn@doi [\aj] {10.1086/301513}, \href
  {http://adsabs.harvard.edu/abs/2000AJ....120.1579Y} {120, 1579}

\bibitem[\protect\citeauthoryear{{Ziparo} et~al.,}{{Ziparo}
  et~al.}{2014}]{zip14}
{Ziparo} F.,  et~al., 2014, \mn@doi [\mnras] {10.1093/mnras/stt1901}, \href
  {https://ui.adsabs.harvard.edu/abs/2014MNRAS.437..458Z} {437, 458}

\bibitem[\protect\citeauthoryear{{von der Linden}, {Wild}, {Kauffmann}, {White}
   \& {Weinmann}}{{von der Linden} et~al.}{2010}]{von10}
{von der Linden} A.,  {Wild} V.,  {Kauffmann} G.,  {White} S.~D.~M.,
  {Weinmann} S.,  2010, \mn@doi [\mnras] {10.1111/j.1365-2966.2010.16375.x},
  \href {http://adsabs.harvard.edu/abs/2010MNRAS.404.1231V} {404, 1231}

\makeatother
\end{thebibliography}

\appendix

\section{Environmental Dependence of Physical Parameters of Galaxies at $z=1.5$ and 2.0} \label{appenA}

\begin{figure*}
	\includegraphics[width=0.85\textwidth]{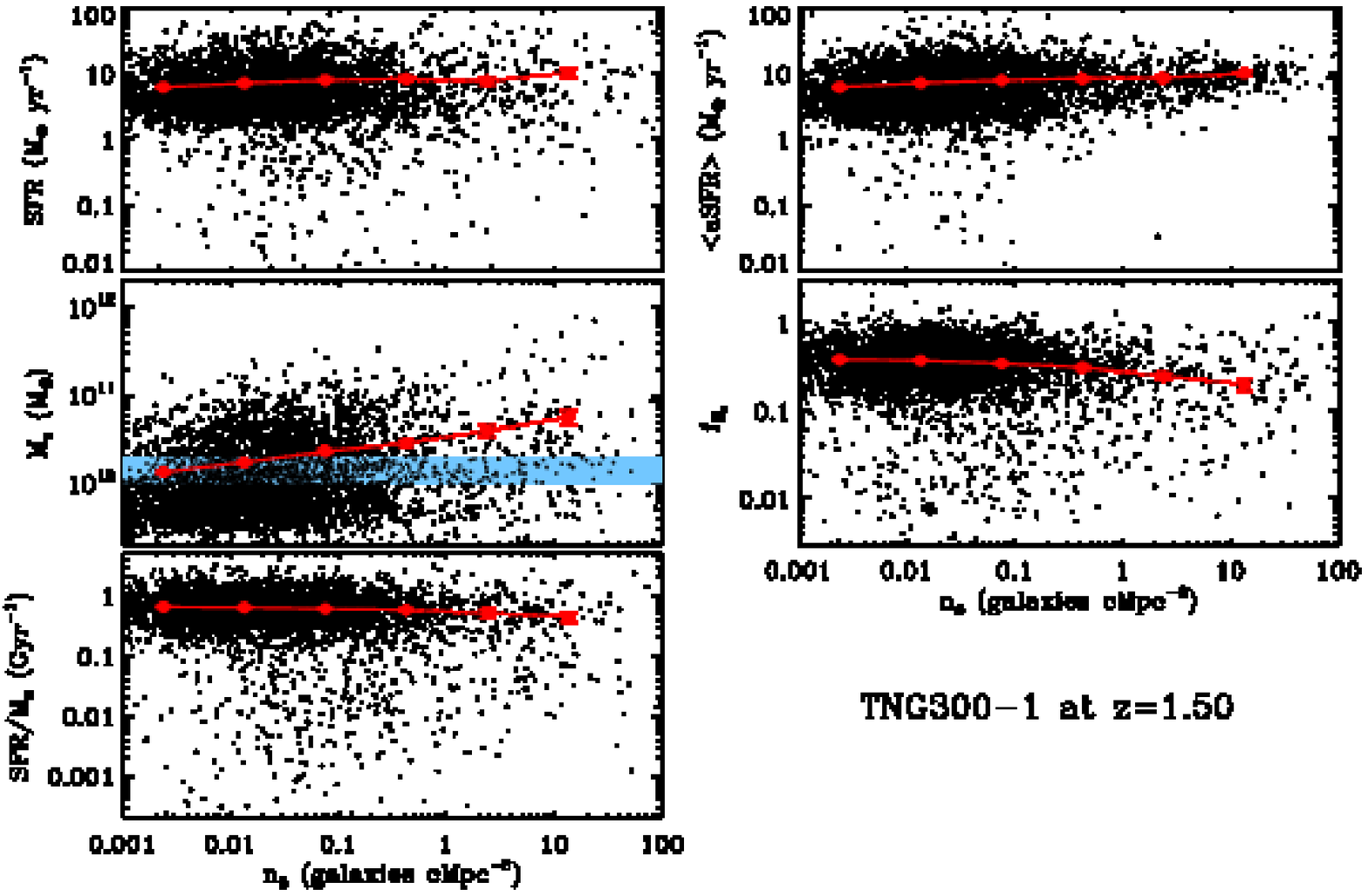}
	\caption{Physical parameters of galaxies in the IllustrisTNG 
		as a function of local density ($n_5$) at $z=1.5$. 
		Black points are individual galaxies, and 
		red points with lines are means of black points.
		The errorbars are derived from the resampling method.
		(top left) star formation rate, 
		(middle left) stellar mass, 
		(bottom left) specific star formation rate,
		(top right) spatially averaged star formation rate, 
		(middle right) mass fraction of molecular hydrogen gas ($f_{\rm H_2} = M_{\rm H_2}/M_{\rm star}$).
		We display only 3\% of the data for clarity.
	}
	\label{z1p5}
\end{figure*}

\begin{figure*}
	\includegraphics[width=0.85\textwidth]{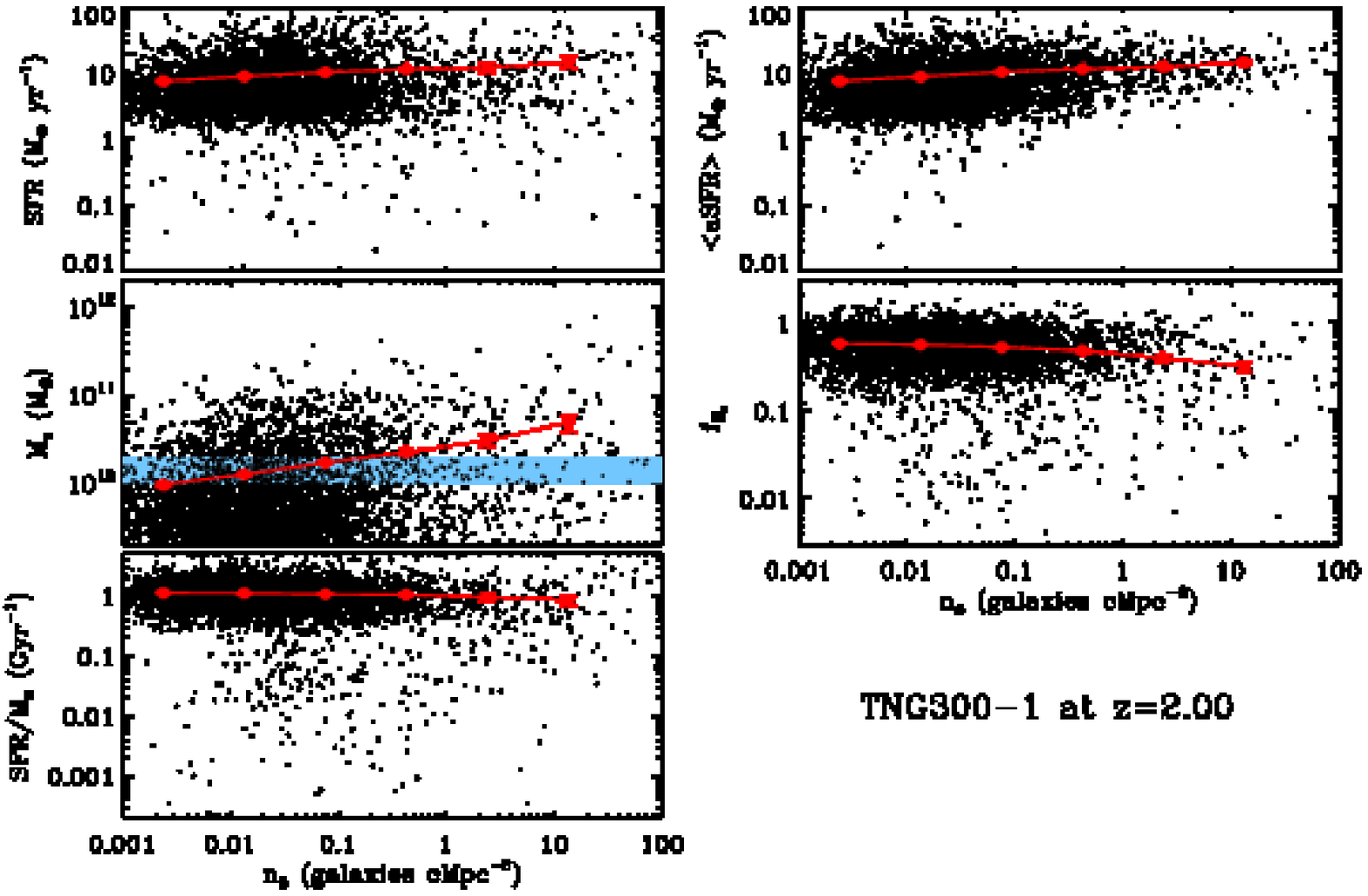}
	\caption{Similar to Fig. \ref{z1p5}, but for $z=2.0$.}
	\label{z2}
\end{figure*}

To better understand how physical parameters of galaxies in the IllustrisTNG
  change with local density at each redshift 
  when the reversal of the SFR-density relation is prominent (i.e. $z>1.0$),
  we show the environmental dependence of physical parameters of galaxies
  at $z=1.5$ and $2.0$ in Figs. \ref{z1p5} and \ref{z2}, respectively.
  
\bsp	

\end{document}